%% file: main.tex
\documentclass[aps,prb,superscriptaddress,tightenlines,nofootinbib,floatfix,longbibliography,notitlepage]{revtex4-2}
\usepackage[utf8]{inputenc}
\usepackage{bm}
\usepackage{amsmath}
\usepackage{amssymb}
\usepackage{tikz}
\usetikzlibrary{arrows}
\usepackage{amsfonts}
\usepackage{graphicx}
\usepackage[english]{babel}
\usepackage{color}
\usepackage{xcolor}
\usepackage[most]{tcolorbox}
\usepackage[left=18mm,right=19mm,top=23mm,bottom=16mm]{geometry}
\usepackage{slashed}
\usepackage{booktabs}
\usetikzlibrary{positioning,calc,arrows.meta} 
\usepackage[
colorlinks=true,
allcolors=blue
]{hyperref}
\usepackage{cleveref}
\usepackage{listings}
\lstnewenvironment{script}{\lstset{
    basicstyle=\ttfamily\small,
    frame=single,
    backgroundcolor=\color{gray!5},
    mathescape=true,
    columns=fullflexible
}}{}

\input{macros}
\input{affiliations}

\begin{document}
\title{Hamiltonian Monte Carlo enhanced by Exact Diagonalization}

\input{authors}
\date{\today}

\begin{abstract} 
	Strongly correlated fermionic systems are of great interest in condensed matter physics and numerical methods are indispensable tools for their study.
	However, existing approaches such as exact diagonalization (ED) and stochastic quantum Monte Carlo methods each suffer from fundamental limitations: ED is hindered by exponential scaling in system size, while Monte Carlo methods are plagued by sign problems and long autocorrelation times.
	These limitations restrict the accessible parameter space and developing algorithms that efficiently alleviate them remains a central challenge in computational physics.
	In this work, we propose a \emph{hybrid} algorithm that combines ED and Hamiltonian Monte Carlo (HMC) to simulate 2D arrays of coupled quantum wires, modeled as interacting fermionic Hubbard chains.
	We demonstrate how our hybrid implementation of HMC, which we dub \algonamedef, outperforms either method alone across several key simulation facets.
	When compared to pure ED, \algoname has a much more favorable computational scaling, which allows us to push simulations to much larger 2D arrays.
	\algoname also greatly alleviates the sign problem and reduces autocorrelation times when compared to pure HMC formulations utilizing either real or imaginary auxiliary fields.  
	Our formalism demonstrates how complementary strengths of seemingly disparate methods can be leveraged to enable feasible simulations in an extended parameter space.
\end{abstract}

\maketitle

\input{section/introduction}

\input{section/method}
\input{section/model}

\input{section/results}

\input{section/summary}

\begin{acknowledgments}
	The code used in this work is publicly available at \cite{HybridHamiltonianMonteCarlo}.
	All data used in the analysis of this study are available from the corresponding author upon reasonable request.
	We are grateful to Maxime Debertolis, Petar Sinilkov, Lin Wang, and Neill Warrington for valuable discussions.
	This work was funded by the Deutsche Forschungsgemeinschaft (DFG, German Research Foundation) as part of the CRC 1639 NuMeriQS – Project number 511713970 and under Germany’s Excellence Strategy – Cluster of Excellence Matter and Light for Quantum Computing (ML4Q) EXC 2004/2 – 390534769.
	We gratefully acknowledge the computing time granted by the JARA Vergabegremium and provided on the JARA Partition part of the supercomputer JURECA at Forschungszentrum Jülich~\cite{jureca-2021}.
\end{acknowledgments}

\bibliography{references}

	\newpage
	\begin{appendix}
		\input{section/appendix}
	\end{appendix}
	
\end{document}

%% file: macros.tex
\usepackage{xspace}
\usepackage{bbm}

\newcommand{\repoURL}{https://github.com/evanberkowitz/latex-base}

\newcommand{\algoname}{H$^2$MC\ }
\newcommand{\algonamedef}{H$^2$MC}

\let\oldexp\exp
\renewcommand{\exp}[1]{\oldexp \left\{ #1\right\}}

\newcommand{\mD}{\mathcal{D}}

\definecolor{cit_gray}{RGB}{35, 38, 54}

\definecolor{lightblueframe}{RGB}{137, 207, 240}
\definecolor{palebackground}{RGB}{235, 245, 255}

\newtcolorbox{rigorbox}{
	colback=palebackground,
	colframe=lightblueframe,
	coltitle=black,
	title=Rigorosity Interlude,
	fonttitle=\bfseries,
	sharp corners,
	before skip=10pt, after skip=10pt,
	boxrule=0.9pt,
	width = \textwidth
}

\newcommand{\secref}[1]{Sec.~\ref{#1}}

\newcommand{\Appref}[1]{Appendix~\ref{#1}}

\newcommand{\figref}[1]{Fig.~\ref{#1}\xspace}

\renewcommand{\eqref}[1]{(\ref{#1})\xspace}
\newcommand{\Eqref}[1]{Eq.~\ref{#1}\xspace}
\newcommand{\R}[1]{Ref.~\cite{#1}}

\newcommand{\Refs}[1]{Refs.~\cite{#1}}

\newcommand{\one}{\ensuremath{\mathbbm{1}}}

\newcommand{\oneover}[1]{\ensuremath{\frac{1}{#1}}}                             %   1/[argument]
\newcommand{\ket}[1]{\ensuremath{\left|\;#1\;\right\rangle}}
\newcommand{\bra}[1]{\ensuremath{\left\langle\;#1\;\right|}}

\let\braket\bracket

\newcommand{\tr}[1]{\ensuremath{\text{tr}\left[#1\right]}}

\let\builtinLaTeX\LaTeX
\def\LaTeX{\builtinLaTeX\xspace}

%% file: affiliations.tex
\providecommand{\orgname}[1]{#1}
\providecommand{\orgaddress}[1]{#1}
\providecommand{\postcode}[1]{#1}
\providecommand{\city}[1]{#1}
\providecommand{\country}[1]{#1}

\newcommand{\hiskp}{
	\orgname{%
		Helmholtz-Institut f\"{u}r Strahlen- und Kernphysik, % (HISKP),
		Universit\"{a}t Bonn,
	}%
	\orgaddress{%
		\postcode{53115} \city{Bonn}, \country{Germany}
	}%
}

\newcommand{\ias}{
	\orgname{%
		Institute for Advanced Simulation 4, % (IAS-4),
		Forschungszentrum J\"{u}lich,
	}%
	\orgaddress{%
		\postcode{52428} \city{J\"{u}lich}, \country{Germany}
	}%
}

\newcommand{\pibonn}{
	\orgname{%
		Physikalisches Institut,
		Universit\"{a}t Bonn
	}%
	\orgaddress{%
		\postcode{53115} \city{Bonn}, \country{Germany}
	}%
}

\newcommand{\bctp}{ 
	\orgname{%
		Bethe Center for Theoretical Physics,
		Universit\"{a}t Bonn
	}%
	\orgaddress{%
		\postcode{53115} \city{Bonn}, \country{Germany}
	}%
}

%% file: authors.tex
\author{Finn L. Temmen}
\thanks{Equal contribution}
\email{f.temmen@fz-juelich.de}
\affiliation{\ias}
\author{Martina Gisti}
\thanks{Equal contribution}
\email{mgisti@uni-bonn.de}
\affiliation{\pibonn}
\affiliation{\bctp}
\author{David J. Luitz}
\email{david.luitz@uni-bonn.de}
\affiliation{\pibonn}
\affiliation{\bctp}
\author{Thomas Luu}
\email{t.luu@fz-juelich.de}
\affiliation{\ias}
\affiliation{\hiskp}
\author{Johann Ostmeyer}
\email{ostmeyer@hiskp.uni-bonn.de}
\affiliation{\bctp}
\affiliation{\hiskp}

%% file: section/introduction.tex
\section{Introduction}
Higher dimensional quantum many-body systems of strongly correlated fermions 
give rise to a wide range of collective phenomena that remain difficult to 
access from first principles, as analytical approaches become intractable in 
regimes where interactions dominate.
Consequently, numerical simulations have become an indispensable tool for 
extracting quantitative insight into such systems, and the development of 
efficient and reliable algorithms remains a central challenge in computational 
physics.

Broadly speaking, common numerical approaches can be divided into methods that 
evaluate many-body quantities directly within the Hilbert space, and stochastic 
methods that reformulate models in terms of a path integral and evaluate it 
stochastically using Monte Carlo techniques. Direct methods operate on an 
explicit representation of the many-body Hilbert space and therefore provide 
numerically exact results.
In particular, exact diagonalization (ED) allows for the computation of 
observables in the thermodynamic equilibrium up to machine precision.
However, the dimension of the Hilbert space grows exponentially with system size, fundamentally limiting the accessible lattice sizes.
This scaling barrier, often referred to as the \emph{curse of dimensionality}, restricts direct approaches to comparatively small systems despite their high accuracy.

Tensor network methods (TN)~\cite{tensor_intro} partially alleviate this constraint by exploiting the entanglement structure of quantum states.
In some cases (especially in 1D), the ground state of a gapped system can be represented with 
polynomial rather than exponential effort.
Unfortunately, excited state properties like the aforementioned gap are typically much harder to extract from TN simulations~\cite{my_tensor_networks}.
Moreover, while one-dimensional matrix product states (MPS)~\cite{Perez-Garcia:2006nqo} have been very successful, simulations of 2D systems are significantly more challenging.
Projected Entangled Pair State (PEPS)~\cite{PEPS_original_bMPS,PEPS_original_2,iPEPS_Corboz2016} are the most natural generalisation of MPS to two dimensions and fermionic versions are readily available~\cite{tensor_fermions}.
Unlike MPS, however, the exact contraction of 2D PEPS, necessary to compute the expectation values, cannot be performed in polynomial time~\cite{Schuch:2007kaz} and all 
practical algorithms must resort to approximations; periodic boundary conditions are then virtually impossible to implement. 
Isometric tensor network states and their fermionic counterparts have been proposed 
precisely to circumvent the contraction difficulties, offering a numerically stable 
subclass of PEPSs that is canonical by construction in 
2D~\cite{Isometric_Zaletel_2020,Fermionic_Dai_2025} and 3D~\cite{Three_Tepaske_2021};
however, the isometric constraint imposes fundamental limitations on the states expressivity.
Alternative 2D ans\"atze such as tree tensor networks (TTNs)~\cite{TTN_Tagliacozzo2009,TTN_Murg2010} and their augmented variants (aTTNs)~\cite{aTTN_cookbook}, which incorporate 
a layer of disentanglers to better capture the entanglement area law in two dimensions, offer a computationally cheaper alternative to PEPS, albeit these represent a smaller 
class of quantum states than PEPS.
Thus, the simulation of large fermionic 2D systems with a small gap remains very challenging if not infeasible for TNs.

Stochastic approaches overcome the scaling limitation by reformulating the evaluation of thermal traces in terms of path integrals, which are then evaluated via stochastic sampling using Monte Carlo integration~\cite{Ostmeyer:2022gwi}.
Specifically, auxiliary fields are introduced by means of a Hubbard-Stratonovich transformation to decouple many-body interactions, and subsequently sampled using a Markov chain Monte Carlo algorithm.
Among these algorithms, the Hamiltonian Monte Carlo (HMC) method~\cite{HMC} (initially hybrid Monte Carlo) enables efficient global updates through molecular dynamics evolution and has long been established as a standard tool in high energy physics and lattice field theory, while increasingly gaining traction in condensed matter applications \cite{Brower:2011av,HMC_MonolayerGraphene,HMC_MonolayerGraphene2,
	HMCGraphene,HMCExtendedHubbardModel,semimetalmott, AF_PhaseTransition,Rodekamp:2024ixu,wang2026hybridmontecarlofractional}.
Nevertheless, practical simulations face two central challenges.
First, complex (or non-positive) statistical weights introduce noise into the estimation of expectation values, giving rise to the notorious sign problem \cite{SignProblemComputationalComplexity,SignProblemManyElectron}, which causes statistical errors to grow.
We remark that formally TNs are not free of a sign problem either~\cite{Chen:2024ggk}, but its severity is typically negligible in practice.
Second, potential barriers and multimodal distributions hinder the efficient exploration of the configuration space, leading to large autocorrelation times and in-practice ergodicity violations that can systematically bias results \cite{Ergodicity_Hubbard}.

A natural strategy to mitigate these difficulties is to hybridize direct and stochastic methods, exploiting the complementary strengths of each method.
With this approach, one can access system sizes well beyond the reach of full ED and avoid the sampling pathologies that afflict purely stochastic formulations, rendering a larger portion of the parameter space accessible.
In this proof of principle work, we employ ED to develop a hybrid HMC (\algonamedef) framework for a two-dimensional system of interacting Hubbard chains.
The considered model consists of one-dimensional Hubbard chains coupled via density-density interactions, forming a genuinely two-dimensional geometry in which inter-chain correlations are driven purely by interactions.
Two dimensional systems are of particular numerical interest, as they are 
particularly challenging for all known methods but seem to lie in reach for 
improvements compared to three dimensional systems.

We demonstrate that the hybrid scheme correctly reproduces exact results and 
yields a significant reduction in both the sign problem and autocorrelation 
times compared to the fully stochastic treatment, enabling access to regimes that 
are simultaneously beyond the reach of full ED and problematic for conventional 
auxiliary field quantum Monte Carlo (AFQMC) approaches.
We further incorporate stochastic trace estimators to improve computational scaling of the algorithm, and probe the two-dimensional character of the model on large systems approaching the computational frontier of the method.

Related hybrid schemes combining MPS with AFQMC methods have 
been explored previously \cite{FusingMPSandMC}, though not in conjunction with 
the HMC algorithm and its associated advantages. We focus here on the same model 
as pursued in Ref. \cite{FusingMPSandMC} for comparability.
More broadly, hybridization strategies have recently been pursued for sign problem reduction via MPS trial wavefunctions in AFQMC for fermionic systems \cite{TrialMPS_AFQMC}.
Hybrid quantum-classical approaches utilizing Lanczos and MPS based techniques have similarly been explored for solving electron-phonon systems \cite{menzler2025hybridquantumclassicalmatrixproductstate}.

The remainder of this paper is organized as follows.
In \secref{sec:method} we briefly summarize the HMC and ED methods hybridized in this work.
In \secref{sec:model} we introduce the two-dimensional model of interacting Hubbard chains and derive the \algoname formalism.
In \secref{sec:results} we present numerical results, assessing the correctness, sampling efficiency, and computational scaling of the hybrid framework across a range of system sizes.
We conclude with a summary and outlook in \secref{sec:summary}.

%% file: section/method.tex
\section{Method}
\label{sec:method}
In this section, we review the two central numerical methods employed throughout this work.
In particular, we summarize the HMC method~\cite{HMC} used for the stochastic estimation of path integrals and the ED approach for the direct evaluation of thermal traces.
For a beginner-friendly introduction to the HMC and its efficient usage we recommend Ref.~\cite{Ostmeyer:2025agg}.

\subsection{Hamiltonian Monte Carlo}
\label{sec:HMC}
In the framework of lattice field theory, a model is formulated in terms of a path integral, where expectation values of physical observables are given by
\begin{equation}
	\label{eq:path_int}
	\langle \mathcal{O}\rangle_S =Z^{-1}\int \mathcal{D}\phi~ \mathcal{O}[\phi] e^{-S[\phi]}
	\quad \text{with} \quad 
	Z= \int \mathcal{D}\phi~  e^{-S[\phi]}, 
\end{equation}
where $\phi_i$, $i=1,\dots,d$ denote the discretized field variables and $\mathcal{D}\phi = \prod_{i=1}^{d} d\phi_i$ a high-dimensional integration measure. 
In most interacting theories, evaluating these integrals analytically is infeasible, and expectation values are instead estimated numerically using Monte Carlo methods.

The basic idea is to interpret $p[\phi] = e^{-S[\phi]}/Z$ as a probability 
distribution of field configurations $\phi$ and to generate an ensemble of 
representative field configurations $\{\phi^{(n)}\}_{n=1}^{N_{\mathrm{cfg}}}$ 
following this distribution. This is of course only possible if $p[\phi]$ is 
non-negative and violations of this condition are common and lead to the so 
called "sign problem".

Then, expectation values defined in \Eqref{eq:path_int} can be expressed by
\begin{equation}
	\langle \mathcal{O}\rangle_S \approx \frac{1}{N_{\mathrm{cfg}}}\sum_{n=1}^{N_{\mathrm{cfg}}} \mathcal{O}[\phi^{(n)}], 
	\quad \text{where} \quad 
	\phi^{(n)} \sim p[\phi].
\end{equation}
This requires an algorithm capable of sampling configurations from $p[\phi]=e^{-S[\phi]}/Z$, such as Markov chain Monte Carlo (MCMC) methods. 
As the name suggests, MCMC methods successively generate a sequence of 
configurations forming a Markov chain, where each state depends only on the 
previous one. This is achieved by introducing a stochastic process with 
transition matrix $\Omega(\phi\rightarrow \phi')$ to move from a configuration $\phi$ to 
another one $\phi'$.
Convergence to a target distribution $p[\phi]$ is guaranteed if the algorithms fulfills two conditions. First, the algorithm has to be ergodic, meaning the algorithm can reach any possible configuration in a finite number of steps.
Second, it fulfills the detailed balance condition
\begin{align}
	\label{eq:detailed_balance}
	p[\phi] \Omega(\phi \rightarrow \phi') = p[\phi'] \Omega(\phi' \rightarrow \phi).
\end{align}

A widely used MCMC algorithm is the HMC method, which introduces a set of conjugate momenta $\pi$ and considers the extended partition function
\begin{align}
	\label{eq:HMC_Z}
	\mathcal{Z} = Z\int \frac{\mathcal{D}\pi}{\mathcal{N}} ~ e^{-\frac{\pi^2}{2}} = \oneover{\mathcal{N}} \int \mathcal{D}\pi\; \mathcal{D}\phi\; e^{-H[\phi,\pi]}, 
\end{align}
where $\mathcal{N}$ is some irrelevant normalization and the Hamiltonian is given by
\begin{align}
	\label{eq:Hamiltonian}
	H[\phi, \pi] = \frac{1}{2}\sum_{i=1}^d \pi_i^2 + S[\phi].
\end{align}
Sampling is then performed in the joint $(\phi,\pi)$ space by evolving Hamilton's equations of motion
\begin{align}
	\label{eq:MD}
	\frac{d\pi}{d\tau} 
	= -\frac{\partial H}{\partial \phi} 
	= -\frac{\partial S}{\partial \phi} 
	\quad\quad \text{and} \quad\quad 
	\frac{d\phi}{d\tau} 
	= \frac{\partial H}{\partial \pi} 
	= \pi
\end{align}
for a fictitious simulation time $\tau$.

For an exact integration, Hamilton's dynamics conserve the energy of the system, ensuring that the configuration maintains the same weight in $\mathcal{Z}$ \eqref{eq:HMC_Z} and thus reproduces the correct target distribution.
In practice, however, the molecular dynamics (MD) equations \eqref{eq:MD} are integrated numerically by discretizing a trajectory of length $t_{\mathrm{MD}}$ into $N_{\mathrm{MD}}$ steps with step size $\epsilon = t_{\mathrm{MD}}/N_{\mathrm{MD}}$.
This inevitably introduces discretization errors that have to be corrected to maintain detailed balance.
Therefore, a newly generated configuration $\tilde{\phi}$ is only accepted according to a Metropolis test with probability
\begin{equation}
	\label{eq:HMC_acceptance}
	p_{\mathrm{HMC}} = \min \left(1, e^{-\Delta H}\right),  
	\quad \text{where} \quad
	\Delta H = H[\tilde{\phi}, \tilde{\pi}] - H[\phi, \pi],
\end{equation}
and rejected otherwise.
If the proposal is accepted, it is appended to the Markov chain, $\phi' = \tilde{\phi}$, otherwise the previous configuration recurs, $\phi' = \phi$.
Throughout this work, we employ the standard leapfrog integrator, which is symplectic and reversible.
Higher order integrators are readily available \cite{OMELYAN2003272,Malezic:2026bds}.

\subsection{Exact diagonalization}
\label{sec:ED}
Fermionic systems pose significant numerical challenges primarily due to the exponential growth of the many-body Hilbert space with increasing system size.
For finite systems, however, the Hamiltonian can be represented exactly in the Fock (occupation-number) basis, allowing fermionic statistics to be treated rigorously.
This approach is commonly referred to exact diagonalization and provides numerically exact results up to machine precision.

Within ED, the Hamiltonian describing a fermionic system is constructed explicitly in the many-body Hilbert space using fermionic creation $c_{i}^{\dagger}$ 
and annihilation $c_{j}^{\vphantom{\dagger}}$ operators, which obey the canonical anticommutation relations
\begin{equation}
	\left \lbrace c_{i}^{\vphantom{\dagger}}, c_{j}^{\dagger} \right\rbrace = \delta_{ij}, \qquad
	\left \lbrace c_{i}^{\vphantom{\dagger}}, c_{j}^{\vphantom{\dagger}} \right\rbrace = \left \lbrace c_{i}^{\dagger}, c_{j}^{\dagger} \right \rbrace = 0,
\end{equation}
and encode the characteristic minus sign that appears when two fermions are exchanged.  The many-body Hilbert space is constructed from the Fock basis, 
where each basis state corresponds to a definite fermionic occupation configuration.
Thermodynamic quantities are obtained from explicit evaluations of thermal traces
\begin{equation}
\label{eq:thermal_trace}
	\langle \mathcal{O} \rangle = Z^{-1}\mathrm{tr}\left(\mathcal{O}e^{-\beta H}\right), \quad \text{where} \quad Z = \mathrm{tr}\left(e^{-\beta H}\right),
\end{equation}
and $\beta$ is the inverse temperature.
In practice, the exact matrix representation of the Hamiltonian is 
constructed by defining operators in the local Hilbert space and iteratively 
building global operators via tensor products.
Fermionic anticommutation relations are preserved by explicitly tracking the fermionic parity, for example through Jordan-Wigner strings or equivalent fermionic sign bookkeeping schemes.
While ED provides numerically exact results for small systems, its applicability is fundamentally limited by the exponential growth of the Hilbert space with system size.
To circumvent this exponential scaling, TN methods \cite{tensor_intro} offer an 
alternative route to estimate thermodynamic quantities by exploiting the entanglement structure of many-body states.

Another strategy to acess larger system sizes is to replace the exact evaluation of traces by stochastic trace estimates~\cite{z2_noise,trace_estimation_overview}.
In this approach, the trace of an operator $\mathcal{O}$ is approximated by averaging expectation values over an ensemble of random states $|\tilde{\psi} \rangle$, 
\begin{equation}
\label{eq:stochastic_trace_estimates}
	\mathrm{tr}(\mathcal{O}) \approx \frac{1}{N_{\text{states}}} \sum_{k=1}^{N_{\text{states}}} \langle \tilde{\psi}_k | \mathcal{O} | \tilde{\psi}_k \rangle,
\end{equation}
where the approximation improves systematically as the number of sampled states $N_{\text{states}}$ increases.

%% file: section/model.tex
\section{Formalism}
\label{sec:model}
In the present section, we begin by introducing the model considered throughout this work and derive a suitable formulation amenable to the treatment with AFQMC methods, specifically with HMC.

\subsection{The model}
In close analogy to \R{FusingMPSandMC}, we consider a system of $L_y$ one-dimensional Hubbard-chains of length $L_x$, coupled by an inter-chain interaction.
The Hamiltonian is given by
\begin{equation}
	\label{eq:full_H_spin}
	\begin{split}
		H 	&= \sum_{j=1}^{L_y} \sum_{i=1}^{L_x}
		\Bigg[
			- \sum_{\sigma=\{\uparrow,\downarrow\}} t_\sigma 
			\left(
				c_{ij\sigma}^\dagger c_{(i+1)j\sigma}^{\vphantom{\dagger}}
				+ \mathrm{h.c.}
			\right) 
			+ U n_{ij\uparrow}n_{ij\downarrow} 
			+ \mu n_{ij} 
			+ V n_{ij} n_{i(j+1)}
		\Bigg], 
	\end{split}
\end{equation}
where $c_{ij\sigma}^{\dagger}$ and $c_{ij\sigma}^{\vphantom{\dagger}}$ create and annihilate an electron with spin $\sigma \in \{\uparrow,\downarrow\}$ at site $i$ on chain $j$, respectively.
Furthermore, we define the number operators $n_{ij\sigma} = c_{ij\sigma}^{\dagger}c_{ij\sigma}^{\vphantom{\dagger}}$ and $n_{ij} = \sum_\sigma n_{ij\sigma}$. 
Throughout this work, we impose periodic boundary conditions in both the $L_x$ and the $L_y$ directions.
Unless stated otherwise, lattice indices labeling sites along a chain run as $i=1,\dots,L_x$, while the chain index runs as $j=1,\dots,L_y$.
In addition to the inter-chain interaction, characterized by the interaction strength $V$, the Hamiltonian contains a nearest neighbor hopping term along the chains with hopping parameters $t_\sigma$, an on-site interaction of strength $U$, and a chemical potential $\mu$. 

\subsection{Decoupling into 1D chains}
Before beginning the derivation, we change to the charge basis by performing the particle-hole transformation 
\begin{equation}
	\label{eq:particle_hole_trafo}
	a_{ij}^\dagger \equiv c_{ij\uparrow}^{\dagger}; \quad
	a_{ij}^{\vphantom{\dagger}} \equiv c_{ij\uparrow}^{\vphantom{\dagger}}; \quad
	b_{ij}^\dagger \equiv c_{ij\downarrow}^{\vphantom{\dagger}}; \quad
	b_{ij}^{\vphantom{\dagger}} \equiv c_{ij\downarrow}^{\dagger}. \quad
\end{equation}
Here, the ladder operators $a_{ij}^{\dagger}$ ($b_{ij}^{\dagger}$) and $a_{ij}^{\vphantom{\dagger}}$ ($b_{ij}^{\vphantom{\dagger}}$) create and annihilate a spin-$\uparrow$ electron (spin-$\downarrow$ electron-hole) at site $i$ on chain $j$, respectively.
Furthermore, we define the local charge operators
\begin{equation}
	\label{eq:number_operators_charge_basis}
	q_{ij} \equiv a_{ij}^\dagger a_{ij}^{\vphantom{\dagger}}, \quad
	\tilde{q}_{ij} \equiv b_{ij}^\dagger b_{ij}^{\vphantom{\dagger}}
	\quad \text{and} \quad
	Q_{ij} \equiv q_{ij} - \tilde{q}_{ij},
\end{equation}
and it follows that $n_{ij\uparrow} = q_{ij}$ and $n_{ij\downarrow} = 1 - \tilde{q}_{ij}$.
Inserting this into the Hamiltonian \eqref{eq:full_H_spin} then yields
\begin{equation}
\label{eq:full_H}
	\begin{split}
		H =& \sum_{i,j}
		\Bigg[
			-\left(
				t_\uparrow a_{ij}^\dagger a_{(i+1)j}^{\vphantom{\dagger}}
				- t_\downarrow b_{ij}^\dagger b_{(i+1)j}^{\vphantom{\dagger}}
				+ \text{h.c.}
			\right) 
			+ (U + \mu + 2V) q_{ij}
			+ (- \mu - 2V)\tilde{q}_{ij} \\
			&- U q_{ij}\tilde{q}_{ij}
			+ V Q_{ij}Q_{i(j+1)}
		\Bigg] + \text{const}, 
	\end{split}
\end{equation}
where we can omit the constant as it cancels out in expectation values.

The first step of the derivation is to decouple the one-dimensional chains by rewriting the inter-chain interaction.
To this end, we complete the square in the $V$ term,
\begin{equation}
	\label{eq:completing_square}
	\begin{split}
		V \sum_{i,j} Q_{ij} Q_{i(j+1)}
		= V \sum_{i,j} (q_{ij} + \tilde{q}_{ij}) - 2V\sum_{i,j}q_{ij} \tilde{q}_{ij} 
		- \frac{V}{2}\sum_{i,j} 
		\left(
			Q_{ij} - Q_{i(j+1)}
		\right)^2
	\end{split}
\end{equation}
and split the Hamiltonian into a purely one-dimensional part acting along the chain $j$, 
\begin{equation}
\label{eq:H1D}
	\begin{split}
		H_{\mathrm{1D}}^{(j)} =& \sum_{i}
		\Bigg[
			-\left(
				t_\uparrow a_{ij}^\dagger a_{(i+1)j}^{\vphantom{\dagger}}
				- t_\downarrow b_{ij}^\dagger b_{(i+1)j}^{\vphantom{\dagger}}
				+ \text{h.c.}
			\right) 
			+ (U + \mu + 3V) q_{ij}
			+ (- \mu - V)\tilde{q}_{ij} 
			- (U+2V) q_{ij}\tilde{q}_{ij} 
		\Bigg].
	\end{split}
\end{equation}
and a quadratic term coupling neighboring chains,
\begin{equation}
	\label{eq:H_V}
	H_V = - \frac{V}{2}\sum_{i,j} 
			\left(
				Q_{ij} - Q_{i(j+1)}
			\right)^2
\end{equation}
such that
\begin{align}
	\label{eq:full_H_rewritten}
	H = \sum_{j} H_{\mathrm{1D}}^{(j)} + H_V.
\end{align}
The additional terms generated by completing the square have been absorbed into the definition of the one-dimensional Hamiltonian \eqref{eq:H1D}. 

\subsection{Path integral formulation}
Next, we discretize the imaginary time interval $[0,\beta]$
into $N_t$ slices of length $\Delta_t = \beta/N_t$ and apply a second-order Suzuki-Trotter 
decomposition on each imaginary time slice
\begin{equation}
\label{eq:Z_trotter_0}
	Z 
	= \mathrm{tr}\left[\prod_{t=1}^{N_t} \left(e^{-\frac{\Delta_t}{2} H_V} e^{-\Delta_t \sum_j H_{\mathrm{1D}}^{(j)}}e^{-\frac{\Delta_t}{2} H_V} + \mathcal{O}(\Delta_t^3)\right)\right].
\end{equation}
Using the cyclicity of the trace, the partition function becomes
\begin{equation}
\label{eq:Z_trotter}
	Z 
	= \mathrm{tr}\left[\prod_{t=1}^{N_t} \left(e^{-\Delta_t \sum_j H_{\mathrm{1D}}^{(j)}}e^{-\Delta_t H_V}\right)\right] + \mathcal{O}(\Delta_t^2),
\end{equation}
where the individual errors in \Eqref{eq:Z_trotter_0} accumulate to an error $\mathcal{O}(\Delta_t^2)$.
Throughout this work, $t=1,\dots,N_t$ labels the imaginary time slices.
In the following, we omit error terms, with the understanding that exact results 
for the partition function and all observables are recovered only in the limit 
$N_t \rightarrow \infty$.

The complete square in \Eqref{eq:H_V} allows us to decouple 
the chains using the 
Hubbard-Stratonovich (HS) transformation\footnote{This step constitutes the main 
difference between the present derivation and that of \R{FusingMPSandMC}. We 
introduce a set of continuous auxiliary fields $\phi$, which enables global 
updates via the HMC algorithm, in contrast to the local updates employed in 
discrete auxiliary field simulations of Hubbard-like models.}
\begin{equation}
\label{eq:HS_trafo}
	\exp{
		\Delta_t \frac{V}{2}
		\left(
			Q_{ij} - Q_{i(j+1)}
		\right)^2
    } \propto \int \limits_{-\infty}^\infty d \phi_{tij}~
	\exp{
		-\frac{1}{2\Delta_t V}\phi_{tij}^2	
		- \phi_{tij} \left(Q_{ij} - Q_{i(j+1)}\right)				
	},
\end{equation}
by introducing an auxiliary field $\phi_{tij} \in \mathbb{R}$ for each lattice 
site $(i,j)$ and time slice $t$.

The transformation in \Eqref{eq:HS_trafo} eliminates all terms coupling different chains (i.e.\ containing products of operators $Q_{ij}Q_{ij'}$), and the Hilbert space factorizes into independent chain contributions. 
The coupling is generated by the auxiliary fields $\phi_{tij}$. We can exploit this tensor product to decompose the trace over the entire Hilbert space from \Eqref{eq:Z_trotter} into a product of traces over the (1D) chain Hilbert spaces:
\begin{align}
	\mathrm{tr} \left[
		\bigotimes_{j=1}^{L_y}\prod_{t}
		\left(
			e^{
				-\Delta_t 
				H_{\mathrm{1D}}^{(j)}
			} 
		e^{
			- \sum_i Q_{ij} \Delta \phi_{tij} 
		}
		\right)
	\right] 
	= \prod_{j=1}^{L_y} \mathrm{tr}_j \left[
			\prod_{t}
			\left(
				e^{
					-\Delta_t 
					H_{\mathrm{1D}}^{(j)}
				} 
				e^{
					- \sum_i Q_{ij} \Delta \phi_{tij} 
				}
			\right)
		\right],
\end{align}
where $\Delta \phi_{tij} \equiv (\phi_{tij} - \phi_{ti(j-1)})$ and $\mathrm{tr}_j$ denotes a trace over the reduced Hilbert space of dimension $4^{L_x}$ associated with the chain $j$.
Combining bosonic and fermionic contributions into the action,
\begin{equation}
	\label{eq:hybrid_action}
	S[\phi] \equiv \sum_{t,i,j} \frac{\phi_{tij}^2}{2\Delta_t V}
				- \sum_{j} \log \mathrm{tr}_j 
				\left[
					\prod_{t}
					\left(
						e^{
							-\Delta_t 
							H_{\mathrm{1D}}^{(j)}
						} 
						e^{
							- \sum_{i} Q_{ij} \Delta \phi_{tij} 
						}
					\right)
				\right],
\end{equation}
the partition function $Z$ and expectation values of observables $\mathcal{O}$ are given by the path integral representation from \Eqref{eq:path_int} with the fermionic part of the action evaluated via the thermal trace in \Eqref{eq:thermal_trace}.

We have now separated the field theoretic description into independent one-dimensionsal, fermionic parts, which we will tackle by ED, and bosonic auxiliary fields $\phi_{tij}$ to be integrated over. This high dimensional integration will be tackled by HMC.

The use of HMC, introduced in \secref{sec:HMC}, additionally requires computation of the gradient of the action \eqref{eq:hybrid_action} with respect to the auxiliary field variables $\phi_{tij}$, which is
\begin{equation}
\label{eq:action_gradient}
	\partial_{\phi_{tij}} S[\phi]
	= \frac{1}{\Delta_t V}\phi_{tij} 
	+ \langle Q_{ij}(t) \rangle
	- \langle Q_{i(j+1)}(t) \rangle,
\end{equation}
where
\begin{equation}
\label{eq:gradient_ev}
	\langle \mathcal{O}_{ij}(t)\rangle 
	\equiv 
	\frac{
		\mathrm{tr}_j
		\left[
			\prod_{t'=1}^{t}
			\left(
				e^{
					-\Delta_t 
					H_{\mathrm{1D}}^{(j)}
				} 
				e^{
					- \sum_{i} Q_{ij} \Delta \phi_{t'ij} 
				}
			\right)
				\mathcal{O}_{ij} \prod_{t'=t+1}^{N_t}
			\left(
				e^{
					-\Delta_t 
					H_{\mathrm{1D}}^{(j)}
				} 
				e^{
					- \sum_{i} Q_{ij} \Delta \phi_{t'ij} 
				}
			\right)
		\right]
	}{
		\mathrm{tr}_j
		\left[
			\prod_{t'=1}^{N_t}
			\left(
				e^{
					-\Delta_t 
					H_{\mathrm{1D}}^{(j)}
				} 
				e^{
					- \sum_{i} Q_{ij} \Delta \phi_{t'ij} 
				}
			\right)
		\right]
	}	.
\end{equation}
Notably, the form of the action and observables is independent of additional terms that could be included in the one-dimensional Hamiltonian \eqref{eq:H1D}, such as next-to-nearest-neighbor hopping, nearest-neighbor interaction, or pairing terms, provided they act within the individual chains.

In practice, the explicit evaluation of thermal traces remains limited by the exponential growth of the Hilbert space, as matrix representations rapidly become prohibitively large.
To enable efficient computations, we avoid forming dense matrix exponentials and instead expand
\begin{equation}
\label{eq:H1D_expansion}
	e^{
		-\Delta_t 
		H_{\mathrm{1D}}^{(j)}
	} = \one 
		- \Delta_t H_{\mathrm{1D}}^{(j)} 
		+ \frac{\Delta_t^2}{2} \left(H_{\mathrm{1D}}^{(j)}\right)^2
		+ \mathcal{O}(\Delta_t^3), 
\end{equation}
leading to sparse matrices due to the locality of $H_{\mathrm{1D}}^{(j)}$.
This drastically reduces memory requirements and computational cost compared to full dense representations.
The truncation error of the expansion of order $\mathcal{O}(\Delta_t^3)$ per time slice adds to the Trotter discretization error introduced in \Eqref{eq:Z_trotter}, and vanishes in the limit $N_t\rightarrow \infty$.

Before concluding this section, we emphasize that the auxiliary field formulation derived above is not unique.
For example, instead of a real HS transformation \eqref{eq:HS_trafo}, one may introduce a purely imaginary auxiliary field $i\phi$.
Such alternative choices lead to formally equivalent representations of the partition function but can differ substantially in their numerical properties.
In preparation for the present work, we explored several decoupling schemes and found the presented formulation to provide the most robust and efficient performance within the parameter regimes considered here.
The implications of this ambiguity are discussed in more detail in \secref{sec:results} and \Appref{sec:pureHMC_formulation}, where we compare different choices in the context of pure HMC formulations of the model.

In summary, we have derived a hybrid formulation in which the auxiliary fields $\phi$ are sampled using Monte Carlo methods, while the fermionic sector is computed via a thermal trace of significantly reduced dimensionality.

%% file: section/results.tex
\section{Results}
\label{sec:results}
In this section, we apply the framework introduced above and assess its performance across different system sizes.
In \secref{sec:benchmark_ED} we first validate the formalism and hybrid approach by studying a small system that permits full ED, enabling a direct benchmark of the computed observables.
We then consider a larger system beyond the reach of ED in \secref{sec:benchmark_HMC} and compare physical observables, autocorrelation times, and the severity of the sign problem with those obtained from a pure HMC implementation of the same model.
Next, in \secref{sec:stochastic_traces} we investigate the use of stochastic trace estimators \eqref{eq:stochastic_trace_estimates} in the computation of gradients and analyze their impact on numerical efficiency and accuracy.
Finally, we discuss computational complexity and push the framework toward the computational frontier in \secref{sec:computational_frontier}, demonstrating the two-dimensional nature of the model by measuring density-density correlation functions across different chains.

\subsection{Benchmarking against exact diagonalization}
\label{sec:benchmark_ED}
We begin by validating the hybrid framework, introduced in \secref{sec:model}, on a system of $L_y = 2$ chains with respective lengths $L_x = 2$, for which full ED remains feasible.
This enables a direct and unbiased comparison of observables and provides a stringent test of both the formalism and its numerical implementation.

We focus on the representative parameter choices $U=3$, $V=1$, $\beta=4$, and $t=1$ at and around half-filling, corresponding to $\mu = -2V -U/2$ on a bipartite lattice.
These parameters were chosen to probe a correlated yet numerically stable regime.
While we refrain from explicitly extrapolating $N_t\rightarrow \infty$ in this benchmark study, we choose two values of $N_t=32,40$ to confirm qualitatively that the simulations converge.
Throughout this work, simulations use the optimal trajectory length \cite{FourierAcceleration}
\begin{equation}
\label{eq:optimal_t_md}
	t_{\mathrm{MD}} = \frac{\pi}{2}\sqrt{\frac{\beta V}{N_t}}, 
\end{equation}
and the number of MD steps $N_{\mathrm{MD}}$ is tuned to obtain acceptance rates of $60-70\%$.
Simulations are initialized from random configurations, thermalized for $N_{\mathrm{therm}}=10^3$ HMC steps, and subsequently $N_{\mathrm{cfg}}=10^4$ configurations are recorded.

\begin{figure}
	\centering
	\includegraphics[width = 0.95\textwidth]{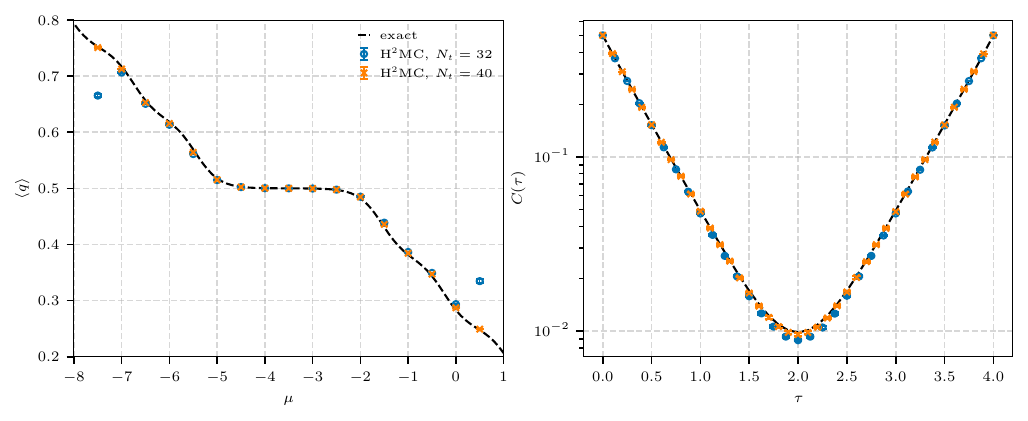}
	\caption{
		Benchmark of the \algoname framework against full ED on a $2\times 2$ lattice at $U=3$, $V=1$, $\beta=4$, and $t=1$ around half-filling, corresponding to $\mu =-3.5$.
		Simulations recorded $N_{\mathrm{cfg}}=10^4$ configurations for $N_t=32$ (blue circles) and $N_t=40$ (orange cross) and are compared to exact continuum results obtained from full ED (black dashed line).
		Left: Average charge density $\langle q\rangle$ \eqref{eq:charge_density} as a function of chemical potential $\mu$. Near half-filling both simulations agree well with the exact result, while deviations grow at large $|\mu|$ for the coarse discretization, reflecting the increasing Suzuki-Trotter error. 
		Right: On-site imaginary-time correlator $C(\tau)$ \eqref{eq:spatial_corr} at half-filling, compared to the exact continuum result. Both values correctly reproduce the correlator accurately across the full imaginary-time extent.}
	\label{fig:placeholder_2x2}
\end{figure}

We first measure the average charge density of spin-$\uparrow$ 
particles\footnote{The density of spin-$\downarrow$ holes follows from $\langle \tilde{q} \rangle = 1 - \langle q \rangle$.} per site, i.e. 
\begin{equation}
\label{eq:charge_density}
	\langle q \rangle \equiv \frac{1}{L_x L_y} \sum_{i,j} \langle q_{ij} \rangle,
\end{equation}
as a function of $\mu$.
The results are depicted in \figref{fig:placeholder_2x2}.
We observe that at and around half-filling, the simulations converge reliably and show excellent agreement with the exact result obtained from ED.
For large absolute values of the chemical potential, the $N_t=32$ simulation exhibits growing deviations from the exact result, reflecting an increasing systematic error due to the Suzuki-Trotter decomposition~\eqref{eq:Z_trotter} and the expansion of the exponential~\eqref{eq:H1D_expansion}.
In contrast, the $N_t=40$ simulation maintains very good agreement, illustrating that these deviations can be systematically reduced by increasing $N_t$.

In addition to the charge density, we also measure the imaginary-time-dependent spatial correlators 
\begin{equation}
\label{eq:spatial_corr}
	C(\tau) \equiv \frac{1}{L_xL_y} \sum_{i,j}
		\langle 
			a_{ij}^{\vphantom{\dagger}}(\tau) a^\dagger_{ij}(0) 	
		\rangle, 
	\quad \text{where} \quad
	\mathcal{O(\tau)} \equiv e^{\tau H} \mathcal{O} e^{-\tau H}. 
\end{equation}
In \figref{fig:placeholder_2x2} the obtained estimates at half-filling from the hybrid approach are compared to the exact continuum result.
Again, we find excellent agreement and observe that both choices of $N_t$ provide accurate results.

Overall, the hybrid method correctly reproduces exact results in the controlled setting of a $2\times 2$ lattice and estimates can be systematically improved by increasing the number of time slices and the number of samples.
Having established its validity in the fully controlled setting, we now turn our attention to system sizes beyond the reach of full ED.

\subsection{Comparing with pure HMC}
\label{sec:benchmark_HMC}
We now compare the performance of the \algoname formalism to two pure HMC formulations of the same model, differing in the choice of HS transformation.
Starting from the action in \Eqref{eq:hybrid_action}, the on-site interaction of strength $U$ is decoupled using a second HS transformation, introducing an additional auxiliary field $\chi$.
As discussed in \secref{sec:model}, this decoupling is not unique as the auxiliary field can be chosen to be purely real, purely imaginary or a mixture thereof \cite{AlgorithmForTheSimulationOfManyElectronSystems,Revisiting_HQMC_for_Hubbard,ConfPhaseTransitionGraphene}.
In this work, we restrict the comparison to the purely real and purely imaginary choices, whose detailed derivations are provided in \Appref{sec:pureHMC_formulation}.
In the following, we refer to these formulations as \emph{real-field pure HMC} ($\chi$) and \emph{imaginary-field pure HMC} ($i\chi$), respectively.

Generally, auxiliary field formulations of interacting fermionic systems are known to suffer from two central limitations.
First, the path-integral weights may become complex (or non-positive), leading to a fermionic sign problem and the degradation of statistical precision.
In such cases, the action is split into a real and imaginary part $S=S_R + iS_I$ and configurations are generated according to the real part of the action $S_R$.
Expectation values are then recovered via reweighting
\begin{equation}
\label{eq:reweighting}
	\langle \mathcal{O} \rangle_S 
		= \frac{ \langle \mathcal{O}e^{-iS_I} \rangle_{S_R}}{\langle e^{-iS_I} \rangle_{S_R}}.
\end{equation}
Second, multimodal probability distributions lead to increased autocorrelation times in HMC simulations as tunneling times increase with growing separation between modes.
Moreover, large autocorrelation times limit the efficient exploration of configuration space and may lead to in-practice ergodicity violations.
Utilizing a real or imaginary HS transformation is often a tradeoff between accepting larger autocorrelation times or sign problem.
Both effects are generally expected to become more severe with increasing coupling parameter, i.e.\@ for the on-site interaction with increasing $U$.

To quantify these two effects, we employ complementary statistical diagnostics. The severity of the sign problem is assessed through the average phase
\begin{equation}
\label{eq:statistical_power}
	\Sigma \equiv |\langle e^{-iS_I} \rangle_{S_R} |.
\end{equation}
General sampling efficiency and the degree of mode trapping can be characterized by the observable-dependent integrated autocorrelation time $\tau_{\mathrm{int},\mathcal{O}}$, which quantifies the extent of correlations within a given time series.
For an observable $\mathcal{O}$, with the measurement on the $i$-th configuration denoted $\mathcal{O}^{(i)}$, it is defined via the autocorrelation function
\begin{equation}
	\label{eq:autocorr_function}
	\Gamma_\mathcal{O}(t) = \bigl\langle \bigl[\mathcal{O}^{(i)} - \langle \mathcal{O} \rangle\bigr] \bigl[\mathcal{O}^{(i+t)} - \langle \mathcal{O} \rangle\bigr]\bigr\rangle, 
\end{equation}
as
\begin{equation}
\label{eq:tint}
	\tau_{\mathrm{int},\mathcal{O}} = \frac{1}{2} \sum_{t=-\infty}^{\infty} \frac{\Gamma_{\mathcal{O}}(t)}{\Gamma_{\mathcal{O}}(0)}.
\end{equation}
For a detailed introduction we refer the reader to Ref.~\cite{Wolff_MonteCarloErrors}.
Together, these two quantities directly determine the effective number of independent samples, which is proportional to $N_{\mathrm{eff,cfg}}\propto \Sigma^2 N_{\mathrm{cfg}} / \tau_{\mathrm{int},\mathcal{O}}$, providing a quantitative basis for comparing the sampling efficiency of the hybrid and conventional approach.

We consider a system of $L_y=6$ chains with length $L_x=4$, corresponding to $24$ lattice sites which is well beyond the reach of full ED.
Choosing $\beta = 1$, $U = 3$, $V = 0.2$, and $t=1$, we tune the chemical potential to half-filling $\mu = -1.9$ and perform both \algoname and pure HMC simulations with $N_t=32$. 
For the \algoname formulation, we first thermalize for $N_{\mathrm{therm}}= 10^3$ trajectories before recording $N_{\mathrm{cfg}}=10^4$ configurations.
For both pure HMC formulations, we use the same thermalization length and record $N_{\mathrm{cfg}}=10^5$ configurations.
All simulations utilize the trajectory length in \Eqref{eq:optimal_t_md} and $N_{\mathrm{MD}}$ is tuned to obtain acceptance rates of $60-70\%$.

\begin{figure}
	\includegraphics[width=0.5\textwidth]{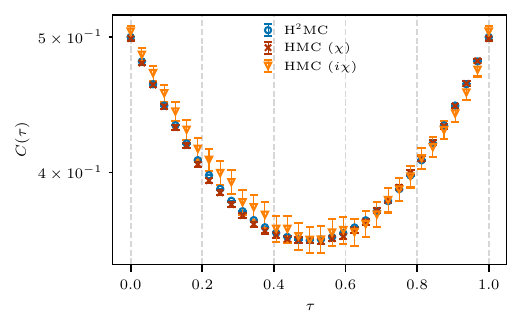}
	\caption{
		On-site imaginary time correlator $C(\tau)$ \eqref{eq:spatial_corr} at half-filling for a $4\times 6$ lattice at $\beta = 1$, $U = 3$, $V = 0.2$, and $t=1$, with chemical potential tuned to half-filling $\mu = -1.9$.
		Simulations compare the \algoname formulation to the real-field $(\chi)$ and imaginary-field $(i\chi)$ pure HMC formulations at $N_t=32$.
		\algoname recorded $N_{\mathrm{cfg}}=10^4$ configurations, while pure HMC generated $N_{\mathrm{cfg}}=10^5$ configurations each.
		All three methods agree within statistical uncertainties, while the imaginary field-pure HMC exhibits visibly enlarged error bars due to a substantial sign problem.
	}
	\label{fig:placeholder_C_up_4x6}
\end{figure}

Results for the spatial correlators \eqref{eq:spatial_corr} are depicted in \figref{fig:placeholder_C_up_4x6}.
All three formulations agree within statistical uncertainties in the estimation of the full temporal extent of the spatial correlators.
However, the imaginary-field pure HMC exhibits a substantial sign problem, with average phase $\Sigma = 0.0748(37)$, leading to visibly enlarged error bars.
In contrast, both the \algoname formulation and the pure HMC with a second real HS transformation did not encounter a sign problem and exhibit comparably small errors.
Nevertheless, since the pure HMC formulation generated an order of magnitude more configurations, this indicates increased autocorrelation times due to the additional HS transformation.

To systematically compare sampling properties, we perform a scan in $U$ at half-filling and fixed $V=0.2$, $\beta = 1$, $t = 1$.
For each simulation, we compute the average phase as well as the integrated autocorrelation time of the spatial correlators \eqref{eq:spatial_corr}.
Since each point of a correlator is associated with its own integrated autocorrelation time, we take the maximal value over all spatial correlators and full time extent
\begin{equation}
\label{eq:max_tint}
	\tau_{\mathrm{int},C} \equiv \max_{\tau ij} \tau_{\mathrm{int},C_{ij}(\tau)}
\end{equation}
to quantify the autocorrelation time of the given simulation.

\begin{figure}
	\includegraphics[width=0.5\textwidth]{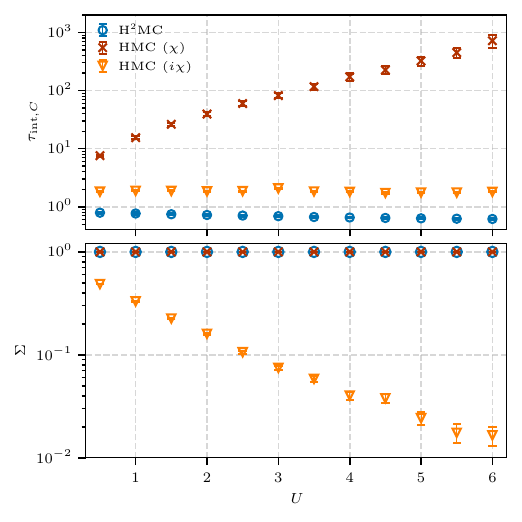}
	\caption{
		Sampling properties of the \algoname, real-field pure HMC ($\chi$), and imaginary-field pure HMC ($i\chi$) formulations as a function of the on-site interaction strength $U$, for a $4\times6$ lattice at half-filling with $\beta = 1$, $V = 0.2$, and $t=1$.
		Top: Maximum integrated autocorrelation time $\tau_{\mathrm{int},C}$ \eqref{eq:max_tint} over all spatial correlator and imaginary-time slices.
		Bottom: Average phase $\Sigma$ \eqref{eq:statistical_power} quantifying the sign problem of the respective simulations.
		The two pure HMC formulations exhibit complementary limitations with increasing $U$; We observe growing autocorrelation times for the real-field and deteriorating average phase for the imaginary-field formulation, while \algoname maintains both a stable short autocorrelation time and an average phase indicating the absence of a sign problem across the full range of $U$ shown.
	}
	\label{fig:placeholder_tint_sigma}
\end{figure}

The results are shown in \figref{fig:placeholder_tint_sigma}.
We observe that the integrated autocorrelation time remains approximately constant for the \algoname and imaginary-field pure HMC, while it shows an exponential increase in the given parameter region for real-field pure HMC.
Conversely, the average phase $\Sigma$ remains stable for \algoname and real-field pure HMC but deteriorates significantly at large $U$ in the imaginary-field pure HMC formulation.
Thus, the two approaches suffer from complementary limitations; either growing autocorrelation times or a worsening sign problem, both leading to reduced statistical precision.
In contrast, by avoiding the second HS transformation, the \algoname method circumvents both issues in this parameter regime.

Naturally, the improved sampling properties must be weighed against the increased computational cost of the \algoname algorithm.
Specifically, in the performed simulations, the time for generating a single configuration in the \algoname formulation amounted to approximately $t_{\mathrm{H}^2\mathrm{MC}}\approx 3.3$s, whereas the generation of a single configuration in the pure HMC formulation took $t_{\mathrm{HMC}}\approx 0.03$s.
Taking both generation time and autocorrelation effects into account, we find that - for our implementation - \algoname is superior for $U\gtrsim 3$ at the current system size, while pure HMC remains more efficient in the weak-coupling regime.

We note that the analysis of this section was extended to a broader range of parameters and the results are shown in detail in \Appref{sec:additional_results}.
In the following we will briefly discuss the main findings of this analysis.

First, we performed the same analysis as presented above for larger values of $V$ up to $V=1$, with qualitatively similar results (see \figref{fig:largerV_Uscan}).
However, the sign problem in the imaginary-field pure HMC formulation worsens rapidly with increasing $V$, rendering that formulation impractical, while real-field pure HMC and \algoname continue to agree.

Similarly, for larger inverse temperatures, the sign problem rapidly renders imaginary-field pure HMC simulations infeasible.
Moreover, increasing $\beta$ while keeping $\Delta_t$ approximately constant also resulted in increasing autocorrelation times in the real-field pure HMC formulation, while \algoname remained largely unaffected in the considered parameter regime (see \figref{fig:beta_scan}).
In addition, at large $\beta$ we encountered numerical instabilities in the na\"ive inversion of the fermion matrix, which is required in the gradients (see \Appref{sec:pureHMC_formulation}).
This is a known issue in AFQMC simulations for which stabilization strategies have been developed \cite{StableDQMC, Bauer:2020stable, AssaadEvertz2008}.
The \algoname approach circumvents this issue entirely, as the gradient \eqref{eq:action_gradient} is computed without requiring matrix inversion.

Finally, we emphasize that similar limitations to those observed with increasing $U$ are expected when increasing $V$ in both the \algoname and pure HMC formalism, since they employ a HS transformation to decouple the inter-chain interaction.
We verified this expectation by measuring $\tau_{\mathrm{int},C}$ as a function of $V$ (see \figref{fig:V_scan}).
We observe that $\tau_{\mathrm{int},C}$ increases drastically around $V\approx 2$, rapidly rendering both \algoname and pure HMC simulations infeasible.
However, from a physical standpoint, the inter-chain interaction strength $V$ is naturally expected to be substantially smaller than the on-site interaction $U$.
Consequently, the regime in which the autocorrelation times become problematic for the $V$ decoupling is unlikely to be reached in physically relevant parameter choices and we do not expect this to constitute a significant practical limitation of the framework.

Overall, the \algoname formulation extends the accessible parameter regime by mitigating the sampling limitations inherent to pure HMC approaches. 
However, despite the improved scaling compared to a full ED calculation, the hybrid algorithm still suffers from the curse of dimensionality in the length of the respective chains.
In the next subsection, we will therefore explore stochastic estimates of the thermal traces encountered in the \algoname formulation to speed up gradient computations.

\subsection{Incorporating stochastic trace estimators}
\label{sec:stochastic_traces}
As discussed in \secref{sec:ED}, computing the exact action gradient becomes computationally intractable as the chains size grows, due to the exponential scaling of the Hilbert space dimension. 
To overcome the dimensional bottleneck, we employ stochastic trace estimators \eqref{eq:stochastic_trace_estimates}, which approximate the full trace using a finite ensemble of randomly sampled quantum states.
This approach is conceptually analogous to the use of pseudofermion fields in AFQMC simulations, where stochastic estimators are similarly introduced to avoid the explicit computation of fermion determinants.
We refer the reader to \Refs{HMC,kennedy2012algorithmsdynamicalfermions} for more details.

We evaluate two candidate sampling strategies for constructing these random states: (i) states whose coefficients are drawn from a complex normal distribution (Gaussian), and (ii) states whose coefficients sampled 
according to a complex Rademacher\footnote{The complex Rademacher distribution is also referred to as $\mathbb{Z}_{4}$ noise distribution.}  distribution~\cite{z2_noise,Stochastic_Hutchinson1989,Random_Iitaka2004,Stochastic_Dong1994}, i.e.\ uniformly from $\{1, i, -1, -i\}$. 
Both distributions satisfy the statistical requirements for an unbiased trace estimator, but may differ in terms of convergence rate, 
numerical stability, and practical efficiency. For a detailed discussion see for instance the appendix of Ref.~\cite{AF_PhaseTransition}.

To quantify the accuracy of each approximation scheme, we compute the root-mean-square error (RMSE) of the stochastically estimated trace 
${\langle {Q_{ij}(t)} \rangle}_{N_{\text{states}}}^{(app)}$ relative to the thermal trace
\begin{equation}
\label{eq:RMSE}
\text{RMSE} = { \left( \frac{1}{N_{\mathrm{cfg}}} \sum_{k=1}^{N_{\mathrm{cfg}}} \left( {\langle {Q_{ij}(t)} \rangle}_{N_{\text{states}}}^{(app)} - {\langle {Q_{ij}(t)} \rangle} \right)_{k}^2 \right) }^{\frac{1}{2}},
\end{equation}
where $N_{\mathrm{cfg}}$ denotes the number of independent configurations used to evaluate the estimator, and $\langle {Q_{ij}(t)} \rangle$ is the corresponding exact trace
of Eq.~\eqref{eq:action_gradient},  serving as reference.
This metric provides a direct, configuration-averaged measure of the deviation of the stochastic estimator from the exact result.

\begin{figure}
     \includegraphics[width=0.5\textwidth, trim={0 0cm 0 0cm}, clip]{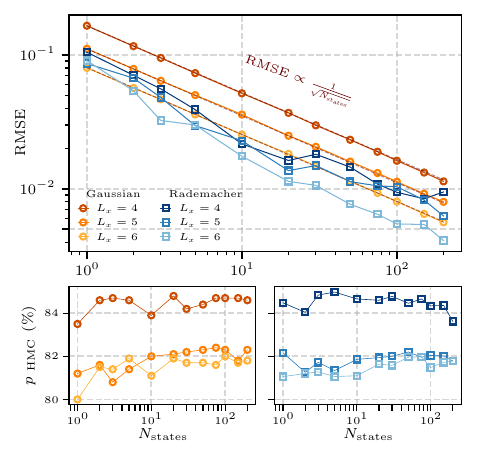}
     \caption{
     	Accuracy and acceptance rate as a function of the number of random states, $N_\text{states}$, sampled from a Gaussian (circles) and Rademacher distribution (squares) for one-dimensional subsystems of length $L_x$.
     	System sizes $L_x= 4,5,6$ are indicated by increasingly lighter shades of orange (Gaussian) and blue (Rademacher).
     	Top: Accuracy of the stochastic estimators quantified by the RMSE \eqref{eq:RMSE} of the stochastically estimated trace relative to the exact results.
     	The Rademacher distribution achieves a consistently lower RMSE than the Gaussian distribution at fixed $N_\text{states}$.
     	Bottom: HMC acceptance rate $p_\text{HMC}$ as a function of $N_\text{states}$ for the Gaussian (left) and Rademacher distribution (right).
     	In both cases, $p_\text{HMC}$ varies only mildly with increasing $N_\text{states}$, confirming the robustness of the HMC dynamics to the stochastic noise of the trace estimator.
	}
	\label{fig:stoch_accuracy}
\end{figure}

In the upper panel of Fig.~\ref{fig:stoch_accuracy}, we report the RMSE as a function of the number of sampled states, $N_{\text{states}}$ for three system sizes ($L_x = 4, 5, 6$). 
Across all system sizes and both sampling distributions, the RMSE decays as $N_{\text{states}}^{-1/2}$, consistent with the standard Monte Carlo convergence rate. This scaling confirms that 
both estimators are unbiased and that their variance decreases predictably with the number of samples - a particularly important property, as it guarantees that accuracy can be systematically 
and controllably improved at a well-defined computational cost. Notably, the Rademacher distribution yields a consistently smaller RMSE than the Gaussian distribution for a given $N_{\text{states}}$, 
suggesting it provides a more efficient basis for trace estimation in this setting.

In the lower panel of Fig.~\ref{fig:stoch_accuracy}, we display the HMC acceptance rate $p_\text{HMC}$ as a function of $N_{\text{states}}$ for both sampling strategies and all system sizes considered. 
A key observation is that $p_\text{HMC}$ remains remarkably stable even at very small values of $N_{\text{states}}$, already reaching its large-$N_{\text{states}}$ plateau for $N_{\text{states}} \lesssim 10$. 
This indicates that the stochastic noise introduced by a crude trace approximation does not significantly destabilize the MD evolution, and that the HMC sampler is robust to the variance of 
the estimator. Furthermore, $p_\text{HMC}$ varies only mildly with system size, confirming that this robustness is not specific to small systems but persists as the Hilbert space dimension grows. 
Taken together, these results establish the Rademacher estimator as the preferred choice: it achieves lower statistical error at fixed computational cost while preserving the efficiency 
of the HMC dynamics.

\subsection{Approaching the computational frontier}
\label{sec:computational_frontier}
We now apply the \algoname framework to larger systems, demonstrating the two-dimensional character of the model.
Before presenting the results, however, we first discuss the computational complexity and memory requirements of the algorithm, as these ultimately determine the accessible system sizes.

We begin by considering the gradient computations \eqref{eq:action_gradient}, and in particular the calculation of the thermal expectation values \eqref{eq:gradient_ev} using the stochastic estimators introduced above.
The computational cost scales linear in both the number of chains $L_y$ and the number of random states $N_{\mathrm{states}}$, so we focus the analysis on the cost for a single chain $j$ and a $\Omega=4^{L_x}$ dimensional random state $\ket{\psi}$.

For a given $\ket{\psi}$, each expectation value requires $N_t$ alternating operations: the matrix-vector products $\ket{\psi'}=\exp{-\sum_i Q_{ij} \Delta \phi_{t'ij}} \ket{\psi}$, which reduces to elementwise multiplication since $Q_{ij}$ is diagonal, followed by a matrix-vector product $\exp{-\Delta_t H_{\mathrm{1D}}^{(j)}} \ket{\psi'}$.
In practice, the expansion \eqref{eq:H1D_expansion} renders the latter matrix exponential sparse.
This results in a total of $(c+1)\Omega$ operations per time slice, where $c$ denotes the average number of nonzero elements per row in the expanded form of the exponential.
For nearest-neighbor hopping and two fermion species with a second-order expansion of the matrix exponential, one estimates $c\approx 5^2$.

The total cost can be reduced substantially by pre-computing and storing the prefix and suffix states
\begin{equation}
\label{eq:prefix_suffix}
	\begin{split}
		\bra{\psi_j(t)} &= 
		\bra{\psi_j} \prod_{t'=1}^{t}\left(
			e^{
				-\Delta_t 
				H_{\mathrm{1D}}^{(j)}
			} 
			e^{
				- \sum_{i} Q_{ij} \Delta \phi_{t'ij} 
			}
		\right), \\
		\ket{\psi_j(t)} &= 
		\prod_{t'=t+1}^{N_t}
		\left(
			e^{
				-\Delta_t 
				H_{\mathrm{1D}}^{(j)}
			} 
			e^{
				- \sum_{i} Q_{ij} \Delta \phi_{t'ij} 
			}
		\right) \ket{\psi_j}
	\end{split}
\end{equation}
for $t=1,\dots,N_t$.
Once these are available, the expectation value at each imaginary time reduces to a vector dot product, and the full set of $N_t$ expectation values is obtained at an additional cost of $N_t\Omega$ operations.
The overall computational complexity is therefore $\mathcal{O}(L_yN_{\mathrm{states}}N_t(2c+2+1)\Omega)$, with memory requirements of $\mathcal{O}(N_t\Omega)$ for storage of the prefix and suffix vectors \eqref{eq:prefix_suffix}.

The evaluation of the partition function entering the action \eqref{eq:hybrid_action} is more demanding, as it is computed exactly rather than stochastically.
It requires $N_t$ sparse matrix-vector products scaling as $c\Omega$, followed by sparse-dense matrix-matrix products scaling as $c\Omega^2$, where repeated multiplications progressively densify the accumulated product. 
This results in a leading order complexity of $\mathcal{O}(L_yN_tc\Omega^2)$, which dominates over the gradient cost and constitutes the bottleneck of the algorithm.
The storage of the accumulated dense matrix further imposes a memory requirement of $\mathcal{O}(\Omega^2)$.

Finally, we note that the computations across chains are embarrassingly parallel, as each chain is treated independently.
Parallelization over $L_y$ thus reduces runtimes up to a factor $L_y$, at the cost of a corresponding increase in total memory consumption.

The above analysis makes clear that the computational frontier of the hybrid algorithm is set by the chain length $L_x$ through the exponential dependence $\Omega=4^{L_x}$, while the number of chains $L_y$ enters only linearly in either runtime or memory when parallelized.
The method is therefore best suited to systems that are extensive in $L_y$ but moderate in $L_x$.
With this in mind, we consider a system of $L_y = 16$ chains of length $L_x=6$, pushing our current implementation toward the computational frontier.
Simulations are performed at $\beta = 1$, $U = 3$, $V = 1$, $t=1$, and half-filling $\mu = -3.5$, following the same thermalization and sampling procedure outlined in \secref{sec:benchmark_HMC}.

To probe that the method captures the two-dimensional nature of the model, we measure the connected density-density correlation function
\begin{equation}
\label{eq:connected_density_density}
	\langle Q_{ij}Q_{i(j+\Delta_j)} \rangle^c = \langle Q_{ij}Q_{i(j+\Delta_j)} \rangle - \langle Q_{ij} \rangle \langle Q_{i(j+\Delta_j)} \rangle,
\end{equation}
between sites on different chains.
Since the chains are not connected by any direct hopping and interact exclusively through the inter-chain density-density coupling \eqref{eq:H_V}, a nonzero inter-chain connected correlator is an unambiguous signature of interaction-induced correlations.
Therefore it provides a direct probe of how the $V$ interaction shapes charge correlations along the $L_y$ dimension.

\begin{figure}
     \includegraphics[width=.5\textwidth]{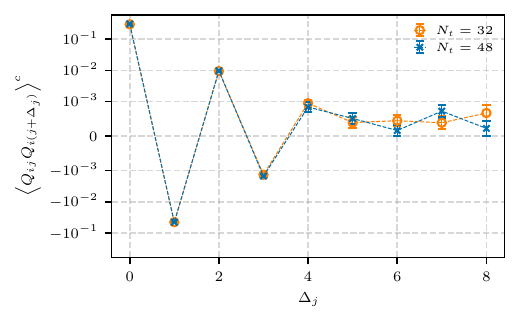}
     \caption{
     	Connected density-density correlation function $\langle Q_{ij}Q_{i'j'} \rangle^c$ \eqref{eq:connected_density_density} as a function of chain separation $\Delta_j$ for a $6\times 16$ lattice at half-filling with $\beta = 1$, $U = 3$, $V = 1$, and $t=1$.
     	Results are shown for two values of $N_t=32,48$, and have been averaged over equivalent separations imposed by the periodic boundary conditions of the system.
     	The correlator alternates in sign with increasing $\Delta_j$, decaying in magnitude with distance.
     	This reflects the repulsive nature of the $V$ interaction coupling neighboring chains.
     }
     \label{fig:6x16}
\end{figure}

The connected density-density correlator as a function of chain separation $\Delta_j$ is depicted in \figref{fig:6x16}.
The correlator has been averaged over equivalent separations imposed by the periodic boundary conditions of the system, leading to fewer than $L_y$ distinct points.
For small chain separations, we observe an oscillating pattern in which the correlator alternates in sign with increasing $\Delta_j$, decaying in magnitude with distance.
This behavior is a direct consequence of the repulsive nature of the inter-chain interaction $V$, where a local presence of charge on one chain favors a depletion on neighboring chains, giving rise to an alternating pattern of correlated charge fluctuations that propagates across the system.
Beyond $\Delta_j=4$, the correlator falls below $10^{-3}$ and the expected alternating patter can no longer be resolved, as the inter-chain correlations become negligibly small at large separations.

%% file: section/summary.tex
\section{Summary and outlook}
\label{sec:summary}

In this work, we have developed and assessed the \algoname framework for a spatially two-dimensional system of interacting Hubbard chains.
By treating the one-dimensional subsystems exactly via ED and sampling the inter-chain degrees of freedom stochastically using HMC, the framework combines the complementary strengths of both approaches.
We benchmarked the method against full ED on small systems, confirming the correctness of the formalism and demonstrating that results can be systematically improved.
For larger systems beyond the reach of full ED, we demonstrated that the hybrid formulation significantly reduces both the sign problem and autocorrelation times compared to pure HMC formulations of the same model.
We further incorporated stochastic trace estimators into the gradient computations required during the MD evolution, showing that they offer a favorable accuracy-to-cost trade-off.
Finally, we discussed the computational complexity of the method and applied it to a large $6\times 16$ system with periodic boundary conditions, where connected density-density correlations between the chains were measured to probe the two-dimensional character of the model.

A similar algorithm has been introduced in Ref.~\cite{FusingMPSandMC}, combining MPS with a standard Metropolis-Hastings algorithm based on a discrete HS transformation.
There, the discrete nature of the auxiliary fields restricts updates to either local Metropolis steps or heuristic proposal schemes. 
In contrast, our use of a continuous HS transformation enables utilization of the HMC algorithm, which performs gradient-informed global updates of the auxiliary field with systematically tunable acceptance rates, leading to a more efficient exploration of the configuration space.
Furthermore, utilizing ED instead of MPS allows for the use of periodic boundary conditions and avoids the systematic error introduced by finite bond dimension truncation.
The primary drawback of this choice is the exponential scaling of ED, which restricts the accessible chain lengths to comparatively small system sizes.

Several directions for addressing this and other limitations naturally suggest themselves for future development.
The most immediate bottleneck of the current implementation is the exact computation of the partition function entering the action, which scales as $\mathcal{O}(L_yN_t\,4^{2L_x})$ and dominates the overall cost.
Extending the stochastic trace estimation approach to the action is therefore a promising route to alleviate this constraint, potentially pushing accessible chain lengths significantly beyond those demonstrated here.

Furthermore, it has recently been shown that augmenting HMC with global multiplicative updates in the radial direction --- so-called \emph{radial updates} --- improves the convergence on non-compact manifolds~\cite{original_radial_update, OstmeyerRadialUpdates} while simultaneously reducing autocorrelation times for some auxiliary field formulations~\cite{FullyErgodicSimulations}.
Since we employ HMC for the stochastic component, incorporating radial updates is a natural extension that we intend to implement in future work. 

A more substantial algorithmic improvement would be to replace the ED component entirely with TN methods, as proposed in Ref.~\cite{FusingMPSandMC}.
Specifically, employing matrix product operators (MPO) to represent the one-dimensional subsystems and time-evolving block decimation (TEBD) for the imaginary-time evolution would alleviate the exponential scaling of ED.
This replacement could dramatically extend the accessible chain lengths while retaining the favorable sampling properties demonstrated in this work.
We emphasize, however, that this substitution is not straightforward, since the 1D chains, on which we calculate the thermal quantities exactly, have periodic boundary conditions, 
which in MPO-based approaches are known to require substantially larger bond dimensions compared to open boundaries.

Finally, extending the model beyond density-density interaction constitutes a natural next step.
Of particular interest are the broader class of arrays of coupled quantum wires as systems exhibiting superconductivity and Majorana physics \cite{Kitaev, TopoSupercondInQuantumWires, TopoSupercondCoupledQuantumWires, RealizationOfKitaev, TwoSiteKitaev, ThreeSiteKitaev}.
They motivate the development of reliable numerical methods capable of treating strongly correlated quasi-one-dimensional systems.
We are confident that our approach to enhance the HMC with ED is generalizable to a vast variety of physical systems as long as they feature relatively weak correlations between small subsystems.
Any potentially very strong interactions within the subsystems can be treated with ED and the remaining dynamics are left to the HMC.
The \algoname framework developed here represents a step in this direction.

%% file: section/appendix.tex
\section{Derivation of the pure HMC formulations}
\label{sec:pureHMC_formulation}
In this section, we summarize the derivation of a pure HMC formulation of the model introduced in \secref{sec:model}.
Starting from the one-dimensional Hamiltonian \eqref{eq:H1D}, our goal is to decouple the on-site interaction of strength $U$ by means of an additional HS transformation.

In contrast to the HS transformation used for the inter-chain interaction $V$, we explicitly derive two equivalent pure HMC formulations by utilizing an ambiguity in the preparation for the HS transformation.
Specifically, completing the square via
\begin{equation}
	-(U + 2V) \sum_{i} q_{ij} \tilde{q}_{ij}
		= \pm \frac{1}{2}(U + 2V) \sum_{i,j} (q_{ij} + \tilde{q}_{ij}) 
		\mp \frac{1}{2}(U + 2V) \sum_{i,j} (q_{ij} \pm \tilde{q}_{ij})^2 
\end{equation}
results in two different formulations depending on the chosen sign convention
\begin{equation}
	H^{\pm}(j) = H_K^{\pm}(j) + H_U^{\pm}(j),
\end{equation}
where we defined
\begin{align}
	H_U^{+}(j) &= 
		- \frac{1}{2}(U + 2V) \sum_{i=1}^{L_x}
		(q_{ij} + \tilde{q}_{ij})^2,
	\quad
	H_U^{-}(j) = 
		\frac{1}{2}(U + 2V) \sum_{i=1}^{L_x}
		(q_{ij} - \tilde{q}_{ij})^2, \\
	H_K^{+}(j) &= \sum_{i=1}^{L_x}
				\Bigg[
					-\left(
						t_\uparrow a_{ij}^\dagger a_{(i+1)j}^{\vphantom{\dagger}}
						- t_\downarrow b_{ij}^\dagger b_{(i+1)j}^{\vphantom{\dagger}}
						+ \text{h.c.}
					\right) 
					+ (\frac{3U}{2} + \mu + 4V) q_{ij}
					+ (\frac{U}{2} - \mu)\tilde{q}_{ij} 
				\Bigg],\\
	H_K^{-}(j) &= \sum_{i=1}^{L_x}
				\Bigg[
					-\left(
						t_\uparrow a_{ij}^\dagger a_{(i+1)j}^{\vphantom{\dagger}}
						- t_\downarrow b_{ij}^\dagger b_{(i+1)j}^{\vphantom{\dagger}}
						+ \text{h.c.}
					\right) 
					+ (\frac{U}{2} + \mu + 2V) (q_{ij} - \tilde{q}_{ij})
				\Bigg].
\end{align}
In complete analogy to the treatment of the $V$ interaction, we apply a second-order Suzuki-Trotter decomposition to separate the interaction contribution from the remaining parts of the Hamiltonian.
Since the on-site interaction consists solely of density operators, it commutes with the auxiliary field contribution originating from the first HS transformation.
Using the cyclicity of the trace, the contributions can therefore be recombined without changing the overall Trotter error, which remains of order $\mathcal{O}(\Delta_t^2)$.
For brevity, we omit writing the error explicitly and reiterate that an extrapolation $N_t\rightarrow\infty$ is required to recover the exact result.

The resulting partition function takes the form
\begin{equation}
	Z = \int \mD \phi~ e^{-S_{\mathrm{B}}[\phi]}
		\prod_{j=1}^{L_y} \mathrm{tr}_j \left[
			\prod_{t=1}^{N_t}
			\left(
				\exp{
					-\Delta_t 
					H_{K}^{\pm}(j)
				} 
				\exp{
					-\Delta_t 
					H_{U}^{\pm}(j)
				} 
				\exp{
					\mp \sum_{i} \Delta \phi_{tij}(q_{ij} - \tilde{q}_{ij})
				}
			\right)
		\right], 
\end{equation}
where $S_{\mathrm{B}}[\phi]\equiv \sum_{t,i,j} \frac{\phi_{tij}^2}{2\Delta_t V}$ denotes the bosonic contribution to the action \eqref{eq:hybrid_action}.
Having isolated the on-site interaction, it can be decoupled via the HS transformation
\begin{equation}
	\exp{ -\Delta_t H_{U}^{\pm}(j)} 
	\propto \int \limits_{-\infty}^\infty d\chi_{tij}
		\exp{
			-\frac{\chi_{tij}^2}{2\Delta_t(U + 2V)}
			- \sigma^\pm \chi_{tij} (q_{ij} \pm \tilde{q}_{ij})
		},
\end{equation}
where $\sigma^\pm \equiv \sqrt{\pm 1}$.
Therefore, for the $(-)$ convention, we obtain a purely imaginary auxiliary field $i\chi$.
As before, the HS transformation comes at the cost of introducing the auxiliary field $\chi_{tij} \in \mathbb{R}$ on every lattice site $(i,j)$ and each imaginary-time slice $t$ that has to be sampled in addition to the $\phi$ field.

Combining both auxiliary field contributions into a single exponential yields
\begin{equation}
\label{eq:Z2_pHMC_der}
	Z = \int \mD \phi\mD \chi~ e^{-S_{\mathrm{B}}[\phi,\chi]}
		\prod_{j=1}^{L_y} \mathrm{tr}_j \left[
			\prod_{t=1}^{N_t}
			\left(
				\exp{
					-\Delta_t 
					H_{\mathrm{1D}}^{\pm}(j)
				} 
				\exp{
					- \sum_{i} 
					\left(
						\Delta \phi_{tij}(q_{ij} - \tilde{q}_{ij})
						+ \sigma^\pm \chi_{tij} (q_{ij} \pm \tilde{q}_{ij})
					\right)
				}
			\right)
		\right], 
\end{equation}
where
\begin{equation}
	S_{\mathrm{B}}[\phi,\chi] = 
		\sum_{t=1\vphantom{j=1}}^{N_t\vphantom{L_y}} 
		\sum_{j=1}^{L_y} 
		\sum_{i=1\vphantom{j=1}}^{L_x\vphantom{L_y}} 
		\left(
			\frac{\phi_{tij}^2}{2\Delta_t V}
			+ \frac{\chi_{tij}^2}{2\Delta_t(U + 2V)}
		\right).
\end{equation}
For compactness and to facilitate the following derivation, these expressions are rewritten as bilinears in the fermionic creation and annihilation operators
\begin{equation}
	\begin{split}
	Z = \int \mD \phi\mD \chi~ e^{-S_{\mathrm{B}}[\phi,\chi]}
		\prod_{j=1}^{L_y} \mathrm{tr}_j \Biggl[
			\prod_{t=1}^{N_t}
			\Biggl(
				&\exp{
					a^{\dagger}_{xj} A^{(\pm)}_{xy} a^{\vphantom{\dagger}}_{yj}
					+ b^{\dagger}_{xj} \bar{A}^{(\pm)}_{xy} b^{\vphantom{\dagger}}_{yj}
				} \\ 
				&\exp{
					a^{\dagger}_{xj} \left(B^{(\pm)}_{tj}[\phi,\chi]\right)_{xy} a^{\vphantom{\dagger}}_{yj}
					+ b^{\dagger}_{xj} \left(\bar{B}^{(\pm)}_{tj}[\phi,\chi]\right)_{xy} b^{\vphantom{\dagger}}_{yj}
				}
			\Biggr)
		\Biggr], 
	\end{split}
\end{equation}
with matrices defined as
\begin{equation}
	\begin{split}
		A^{(+)}_{xy} &= 
			\Delta_t t_\uparrow \delta_{\langle x,y\rangle} 
			- \Delta_t 
				\left(
					\frac{3U}{2} + \mu + 4V  
				\right) \delta_{xy}\\
		\bar{A}^{(+)}_{xy} &= 
			- \Delta_t t_\downarrow \delta_{\langle x,y\rangle}
			- \Delta_t 
				\left(
					\frac{U}{2} - \mu  
				\right) \delta_{xy}\\
		A^{(-)}_{xy} &= 
			\Delta_t t_\uparrow \delta_{\langle x,y\rangle} 
			- \Delta_t 
				\left(
					\frac{U}{2} + \mu + 2V  
				\right) \delta_{xy}\\
		\bar{A}^{(-)}_{xy} &= 
			- \Delta_t t_\downarrow \delta_{\langle x,y\rangle}
			+ \Delta_t 
				\left(
					\frac{U}{2} + \mu + 2V  
				\right) \delta_{xy}\\
		\left(B^{(\pm)}_{tj}[\phi,\chi]\right)_{xy} &= 
			\left(
				- \Delta \phi_{txj} - \sigma^\pm \chi_{txj}
			\right) \delta_{xy}\\
		\left(\bar{B}^{(\pm)}_{tj}[\phi,\chi]\right)_{xy} &= 
			\left(
				\Delta \phi_{txj} \mp \sigma^\pm \chi_{txj}
			\right) \delta_{xy}					
	\end{split}. 
\end{equation}
These matrix definitions will be used throughout the remainder of this section and appear in the final expressions.
The derivation can be straightforwardly generalized to include additional terms such as next-to-nearest neighbor hopping.

For each chain $j$, the thermal trace is evaluated by inserting $2N_t$ sets of fermionic coherent states between successive exponential operators in \Eqref{eq:Z2_pHMC_der} using the completeness relation
\begin{equation}
	\one = \int \mD(\bar{\psi}_{lj}\psi_{lj}\bar{\eta}_{lj}\eta_{lj})
	\exp{
		-\sum_{i} 
		(
			\bar{\psi}_{lij}\psi_{lij}
			+\bar{\eta}_{lij}\eta_{lij}
		)
	}
	\ket{
		\psi_{lj},\eta_{lj}
	}
	\bra{
		\psi_{lj},\eta_{lj}
	},
\end{equation}
where $\mD(\bar{\psi}_{lj}\psi_{lj}\bar{\eta}_{lj}\eta_{lj}) = \prod_{i=1}^{L_x}d\bar{\psi}_{lij}d\psi_{lij}d\bar{\eta}_{lij}d\eta_{lij}$ and
\begin{equation}
	\ket{\psi_{lj},\eta_{lj}} = \exp{
		\sum_{i}
		(
			\psi_{lij} a_{ij}^\dagger
			+\eta_{lij} b_{ij}^\dagger
		)
	} \ket{0}
	\quad \text{and}\quad 
	\bra{\psi_{lj},\eta_{lj}} = \bra{0}
	\exp{
		-\sum_{i}
		(
			a_{ij}\bar{\psi}_{lij} 
			+b_{ij}\bar{\eta}_{lij} 
		)
	}.
\end{equation}
Using the definition of the thermal trace $\mathrm{tr}\left(A\right)=\sum_n \bra{n}A\ket{n}$ together with the identities 
\begin{align}
	\braket{n}{\psi,\eta} \braket{\psi,\eta}{n}
	&= \braket{-\psi,-\eta}{n} \braket{n}{\psi,\eta}
		\quad \text{and} \quad \\
	\bra{\psi} 
		\exp{
			\sum_{x,y} c^\dagger_{x} A_{xy} c_{y}
		} 
	\ket{\psi} 
	&= \exp{
		\sum_{x,y} \bar{\psi}_x \left(e^A\right)_{xy} \psi_y
		},
\end{align}
fermionic ladder operators can be replaced by Grassmann variables, yielding
\begin{equation}
\label{eq:Z3_pHMC_der}
	\begin{split}
	Z = &\int \mD \phi\mD \chi\mD(\bar{\psi}\psi\bar{\eta}\eta)~ e^{-S_{\mathrm{B}}[\phi,\chi]} \exp{-\sum_{l=1}^{2N_t}\sum_{i,j}\left(\bar{\psi}_{lij}\psi_{lij} + \bar{\eta}_{lij}\eta_{lij}\right)} \\
		\prod_{j} 
				&\exp{
					\sum_{t,i}
						\left(
							\bar{\psi}_{(2t)xj} \left(e^{A^{(\pm)}}\right)_{xy} \psi_{(2t+1)yj}
							+ \mathcal{B}_t \bar{\psi}_{(2t+1)xj} \left(e^{B_{tj}^{(\pm)}[\phi,\chi]}\right)_{xy} \psi_{(2t+2)yj}
						\right)
				} \\
		\prod_{j} 
				&\exp{
					\sum_{t,i}
						\left(
							+ \bar{\eta}_{(2t)xj} \left(e^{\bar{A}^{(\pm)}}\right)_{xy} \eta_{(2t+1)yj}
							+ \mathcal{B}_t \bar{\eta}_{(2t+1)xj} \left(e^{\bar{B}_{tj}^{(\pm)}[\phi,\chi]}\right)_{xy} \eta^{\vphantom{\dagger}}_{(2t+2)yj}
						\right)
				}.
	\end{split}
\end{equation}
Here $\mathcal{B}_t$ encodes the anti-periodic boundary conditions in the imaginary-time direction and is defined as $\mathcal{B}_{t}=+1$, for $t<N_t$ and $\mathcal{B}_{N_t}=-1$.
For a detailed introduction to Grassmann variables and fermionic coherent states we refer the reader to Ref.~\cite{Negele:1988vy}.

The Grassmann variables can now be integrated out, producing fermion determinants of dimension $(2N_tL_x)$ for each fermion species and chain $j$.
In practice, it is computationally advantageous to first employ Grassmann $\delta$-functions (see Ref.~\cite{ZinnJustin}) to reduce the dimensionality of the fermionic matrices, resulting in
\begin{equation}
	Z = \int \mD \phi\mD \chi~ e^{-S^{\vphantom{(\pm)}}_{\mathrm{B}}[\phi,\chi]-S^{(\pm)}_{\mathrm{F}}[\phi,\chi]}, 
\end{equation}
with the fermionic action
\begin{equation}
	S^{(\pm)}_{\mathrm{F}}[\phi,\chi] \equiv -\sum_{j=1}^{L_y} \left( \log\det M_j^{(\pm)}[\phi,\chi] + \log\det \bar{M}^{(\pm)}_j[\phi,\chi] \right)
\end{equation}
where 
\begin{equation}
	\begin{split}
		M_j^{(\pm)}[\phi,\chi] &= \one + \prod_{t=1}^{N_t} \left( e^{A^{(\pm)}} e^{B_{tj}^{(\pm)}[\phi,\chi]} \right), \\
		\bar{M}^{(\pm)}_j[\phi,\chi] &= \one + \prod_{t=1}^{N_t} \left( e^{\bar{A}^{(\pm)}} e^{\bar{B}_{tj}^{(\pm)}[\phi,\chi]} \right)
	\end{split}
\end{equation}
are matrices of dimension $L_x$.
Utilization of HMC requires the computation of force terms, which involve derivatives of the logarithm of the fermion matrix determinant.
For a general fermion matrix $M[\phi]$, using Jacobi's formula yields
\begin{equation}
\label{eq:pureHMC_gradient}
	\frac{\partial}{\partial \phi} \log \det M[\phi] = \tr{M^{-1}[\phi]\frac{\partial}{\partial \phi }M[\phi]},
\end{equation}
requiring the explicit computation of the inverse fermion matrix $M^{-1}[\phi]$.

This completes the construction of the real-field $(+)$ and imaginary-field $(-)$ pure HMC formulations employed in \secref{sec:benchmark_HMC}.

\section{Additional results}
\label{sec:additional_results}
In this appendix, we collect additional numerical results complementing the comparison between \algoname and the pure HMC formulations of \secref{sec:benchmark_HMC}.
Specifically, we extend the analysis performed there to a broader range of parameters, examining the dependence of the integrated autocorrelation $\tau_{\mathrm{int},C}$ \eqref{eq:max_tint} and average phase $\Sigma$ \eqref{eq:statistical_power} on the inter-chain interaction strength $V$, the inverse temperature $\beta$, and the on-site interaction $U$ at larger $V$.
Throughout, the system size is kept at a $4\times 6$ lattice.

\begin{figure}
	\includegraphics[width=0.5\textwidth]{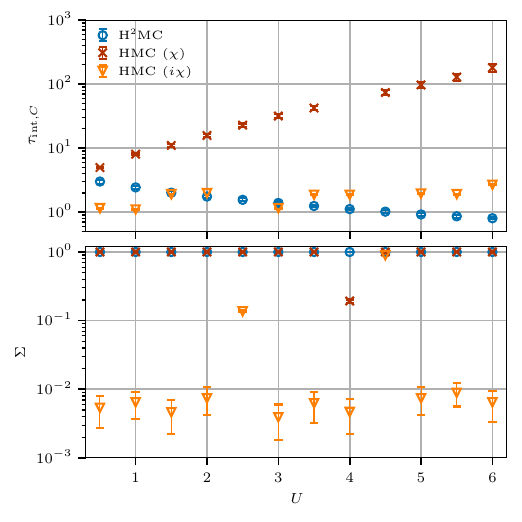}
	\caption{
		Sampling properties of the \algoname, real-field pure HMC ($\chi$), and imaginary-field pure HMC ($i\chi$) formulations as a function of the on-site interaction strength $U$, for a $4\times6$ lattice at half-filling with $\beta = 1$, $V = 1$, and $t=1$.
		Top: Maximum integrated autocorrelation time $\tau_{\mathrm{int},C}$ \eqref{eq:max_tint} over all spatial correlator and imaginary-time slices.
		Bottom: Average phase $\Sigma$ \eqref{eq:statistical_power} quantifying the sign problem of the respective simulations.
		The two pure HMC formulations exhibit complementary limitations with increasing $U$; We observe growing autocorrelation times for the real-field and a prohibitively small average phase for the imaginary-field formulation, while \algoname maintains both a stable short autocorrelation time and an average phase indicating the absence of a sign problem across the full range of $U$ shown.
	}
	\label{fig:largerV_Uscan}
\end{figure}

We first repeat the $U$ scan of \secref{sec:benchmark_HMC} at a larger inter-chain interaction strength $V=1$, probing whether the qualitative picture observed at smaller $V$ persists in a more strongly coupled regime.
The results are shown in \figref{fig:largerV_Uscan}.
We again observe increasing $\tau_{\mathrm{int},C}$ for the real-field pure HMC and a diminishing average phase $\Sigma$ in the imaginary-field pure HMC formulation.
In contrast, \algoname retains small autocorrelation time and does not exhibit a sign problem, underlining the results of \secref{sec:benchmark_HMC}.
Furthermore, we observe several outlier pure HMC simulations, where numerical instabilities in the inversion of the fermion matrix resulted in near-zero acceptance rates.

\begin{figure}
	\includegraphics[width=0.5\textwidth]{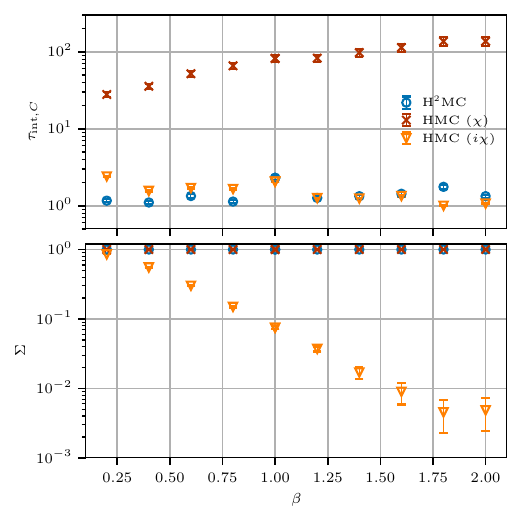}
	\caption{
		Sampling properties of the \algoname, real-field pure HMC ($\chi$), and imaginary-field pure HMC ($i\chi$) formulations as a function of the inverse temperature $\beta$ at approximately constant discretization $\Delta_t\approx 1/32$, for a $4\times6$ lattice at half-filling with $U=3$, $V = 0.2$, and $t=1$.
		Top: Maximum integrated autocorrelation time $\tau_{\mathrm{int},C}$ \eqref{eq:max_tint} over all spatial correlator and imaginary-time slices.
		Bottom: Average phase $\Sigma$ \eqref{eq:statistical_power} quantifying the sign problem of the respective simulations.
		The two pure HMC formulations exhibit complementary limitations with increasing $\beta$; We observe growing autocorrelation times for the real-field and a deteriorating average phase for the imaginary-field formulation, while \algoname maintains both a stable short autocorrelation time and an average phase indicating the absence of a sign problem across the full range of $\beta$ shown.
	}
	\label{fig:beta_scan}
\end{figure}

Next, we extend the analysis to larger inverse temperatures, performing simulations up to $\beta = 2$ for fixed $U=3$, $V=0.2$, and $t=1$ at half-filling.
The Trotter discretization $\Delta_t$ is held approximately constant by adjusting $N_t$ to ensure that any observed trends are not contaminated by increasing discretization errors.
The results are shown in \figref{fig:beta_scan}.
The $\beta$ dependence exhibits a qualitatively similar behavior to the $U$ scan, where pure HMC formulations suffer from complementary limitations, while \algoname does not exhibit a sign problem and maintains small autocorrelation times.

\begin{figure}
	\includegraphics[width=0.5\textwidth]{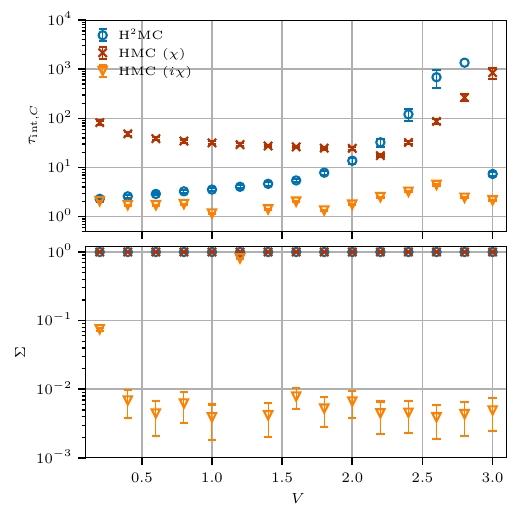}
	\caption{
		Sampling properties of the \algoname, real-field pure HMC ($\chi$), and imaginary-field pure HMC ($i\chi$) formulations as a function of the inter-chain interaction strength $V$, for a $4\times6$ lattice at half-filling with $\beta = 1$, $U = 3$, and $t=1$.
		Top: Maximum integrated autocorrelation time $\tau_{\mathrm{int},C}$ \eqref{eq:max_tint} over all spatial correlator and imaginary-time slices.
		Bottom: Average phase $\Sigma$ \eqref{eq:statistical_power} quantifying the sign problem of the respective simulations.
		Both \algoname and real-field pure HMC formulations exhibit significantly increasing autocorrelation times at large $V$, while the imaginary-field pure HMC develops a prohibitively small average phase already at small $V$.
		Real-field pure HMC and \algoname show an average phase indicating the absence of a sign problem across the entire range of $V$ shown.
		We also observe an outlier simulation for the imaginary-field pure HMC simulation at $V=1.2$, where numerical instabilities resulted in a near-zero acceptance rate.
	}
	\label{fig:V_scan}
\end{figure}

Finally, we consider a scan in the inter-chain interaction strength $V$ at fixed $U=3$, $\beta = 1$, and $t=1$.
The results for $\tau_{\mathrm{int},C}$ and $\Sigma$ as a function of $V$ are shown in \figref{fig:V_scan}.
We observe that both \algoname and real-field pure HMC exhibit significantly increasing autocorrelation times with growing $V$, consistent with the expectation that the HS transformation decoupling the inter-chain interaction introduces similar sampling difficulties as the $U$ decoupling.
The autocorrelation times increase drastically around $V\approx 2$, rapidly rendering both formulations infeasible in this regime.
In fact, for the \algoname simulation at $V=3$, we observe an in-practice ergodicity violation that leads to a heavily underestimated autocorrelation time.
The imaginary-field pure HMC formulation develops a prohibitively small average phase already at small $V$ and exhibits an outlier simulation at $V=1.2$, where numerical instabilities resulted in a near-zero acceptance rate.
